\documentclass[cits]{PoS}

\usepackage{graphics}
\usepackage{epsfig}

\def\apj{ApJ}%
\def\apjl{ApJ}%
          \def\apjs{ApJS}%
\def\aap{A\&A}%
\def\mnras{MNRAS}%
\title{Fermi Observations of Blazars: Implications for Gamma-ray Production}

\ShortTitle{Fermi Observations of Blazars}

\author{\speaker{Juri Poutanen} \\ % \thanks{A footnote may follow.}\\
Astronomy Division, Department of Physics, PO Box 3000, FI-90014 University of Oulu, Finland \\
        E-mail: \email{juri.poutanen@oulu.fi}}

\author{Boris E.  Stern\\
        Institute for Nuclear Research,  Moscow, Russia \\ 
        Astro Space Center, Lebedev Physical Institute, Moscow, Russia\\
        E-mail: \email{stern.boris@gmail.com}}

\abstract{The brightest blazars detected by the {\em Fermi Gamma-ray Space Telescope} Large Area Telescope ({\em Fermi}/LAT)
show significant breaks in their spectra at a few GeV.  
The sharpness and the position of the breaks can be well reproduced by absorption of $\gamma$-rays via 
photon--photon pair production on  He\,{\sc ii}  and  H\,{\sc i}  Lyman recombination continua  (LyC) 
produced in the broad-line region  (BLR).  
Using 138 weeks of LAT observations of the brightest GeV blazar 3C~454.3 we find 
a power-law dependence of the peak energy on flux 
and discover anti-correlation between  flux and 
the column density of the \mbox{He\,{\sc ii}}  LyC which is responsible for absorption of the $>2.5$ GeV photons in this object. 
The strength and the variability of the absorption implies the location of the $\gamma$-ray emitting zone 
close to the boundary of the high-ionization part of the BLR 
and moving away from the black hole when the flux increases.
A combination of the GeV breaks with the detection of a few powerful blazars in the TeV range puts 
strong constraints on the BLR size.  
Additional spectral breaks  at $\sim$100 and $\sim$400 GeV
due to absorption by the Balmer and Paschen lines could be detected by the Cherenkov Telescope Array.
}

\FullConference{AGN Physics in the CTA Era - AGN2011,\\
		May 16-17, 2011\\
		Toulouse, France}

\begin{document}

\section{Fermi spectra of blazars}

\begin{figure}[h]
\centerline{\epsfig{file=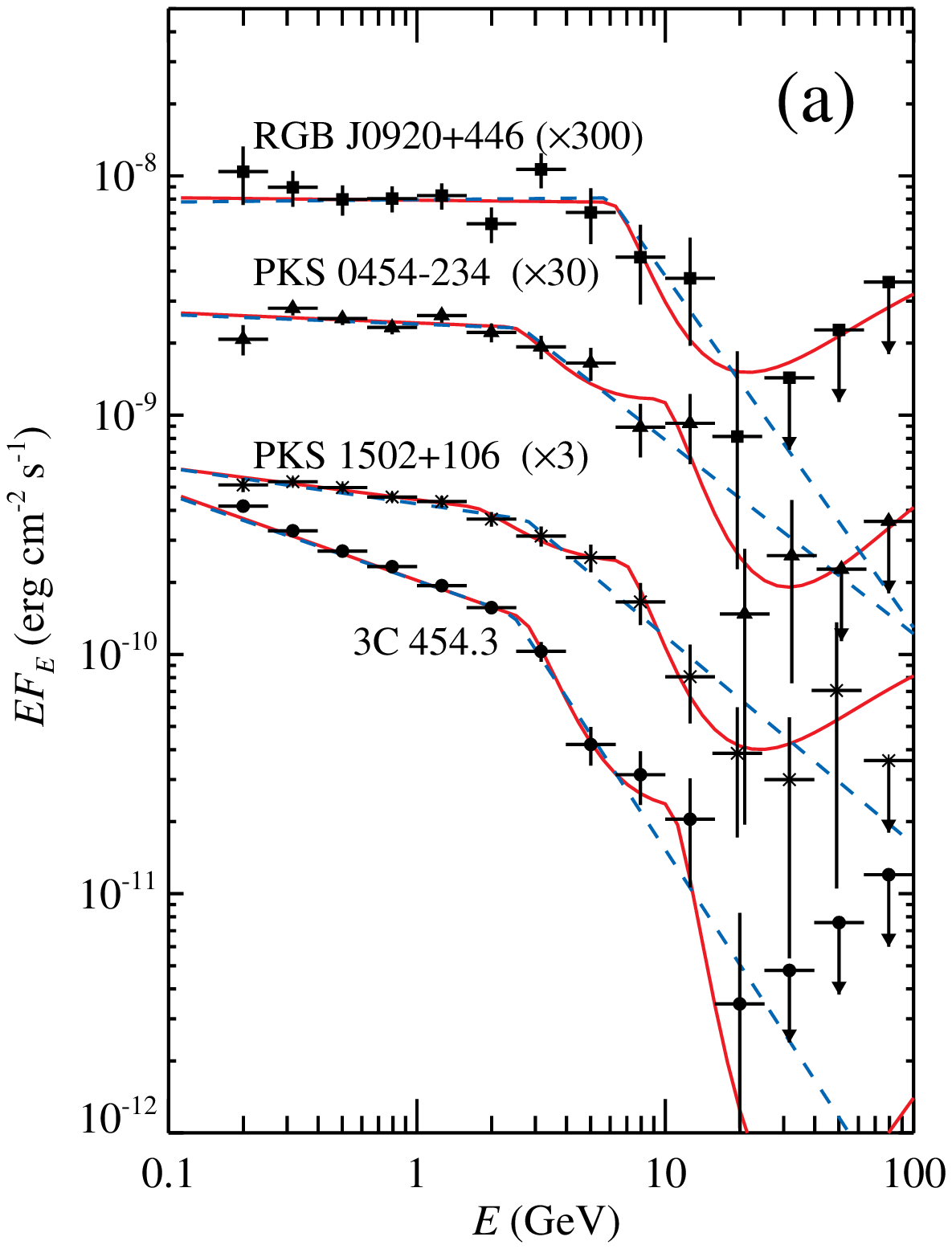,width=6cm} \hspace{1cm}
\epsfig{file=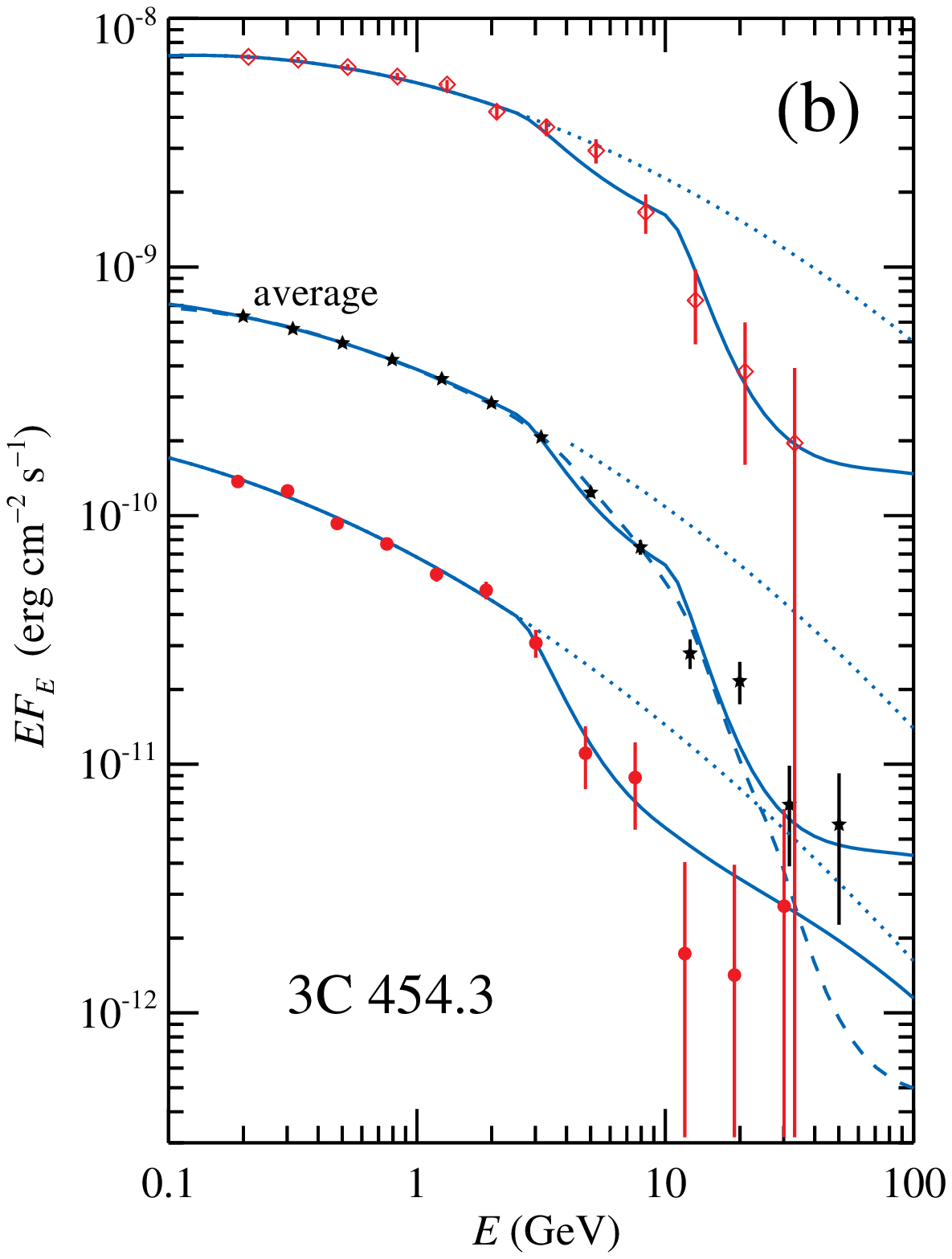,width=6cm} } 
\caption{(a) Spectral energy distribution of a few blazars as observed by {\em Fermi}/LAT 
during first 6 months  of its operation.  
The  best-fit broken power law and  a power law with the absorption  
by the \mbox{He\,{\sc ii}} and H LyC (double-absorption) are shown by  the dashed and solid lines, respectively.  From \cite{PS10}.
(b) Spectral energy distribution of 3C~454.3 at low and high fluxes (intervals B and H, see Fig. \protect\ref{fig:lc}) 
as well as averaged over the whole observation period of 2.5 years. 
Here the underlying continuum is modeled as the lognormal function (shown by the dotted curve). 
The models modified by double-absorption are shown by solid curves, while the dashed curve is for absorption by BLR 
at a fixed ionization parameter $\log \xi=2.5$.  From \cite{SP11}.
}\label{fig:spectra}
\end{figure}

The {\em Fermi}/LAT has detected a few hundred blazars in the 100 MeV--100 GeV range \cite{Abdo09_AGN}. 
The high photon statistics from the brightest sources allows also to study their spectra with unprecedented accuracy. 
These studies have shown that the spectra of high-luminosity sources, flat-spectra radio quasars (FSRQs) and low-energy synchrotron peaked BL Lac objects, are much better described by a broken power law than by a simple powerlaw or any smoothly curved models \cite{Abdo09_3C454.3,Abdo10_blazars} (see Fig. \ref{fig:spectra}a). 
The works dedicated to 3C~454.3   \cite{Fermi10_3C454,Fermi11_3C454} 
found the spectral hardness--flux correlation at a nearly constant break energy. 
 
The observed spectral breaks are too sharp to be associated with the cooling or the Klein-Nishina effects \cite{Fermi10_3C454}. External Compton scattering on a  truncated electron spectrum  \cite{Abdo09_3C454.3} also cannot possibly 
reproduce the break sharpness. 
Na\"{i}vely, one would think that the break energies, 
mostly lying below 10 GeV (as measured in the object frame, see Fig. \ref{fig:spectra}a),  
are too small  to be consistent with the  $\gamma$-ray absorption due to photon--photon ($\gamma\gamma$) pair production in the broad-line region (BLR), as the strongest BLR line,  
Ly$\alpha$, absorbs $\gamma$-rays starting only from 25.6 GeV \cite{Abdo10_blazars}.  
However, the BLR contains not only hydrogen, but also e.g. helium, and emits not only lines, 
but also strong recombination continua.
The observed break energies  \cite{Abdo10_blazars,PS10} 
favour  the \mbox{He\,{\sc ii}} and \mbox{H\,{\sc i}} Lyman recombination continua (LyC) as 
the main opacity source. 
A relatively high opacity in the \mbox{He\,{\sc ii}} LyC observed in a number of bright blazars 
implies the location of the $\gamma$-ray emitting region within the highly ionized inner part of the BLR \cite{PS10}. 

\section{Broad-line region and GeV opacity}

It is well known  that BLRs around quasars emit lines associated with different ionization stages, 
with the broader high-ionization lines being produced closer 
to the central black hole   \cite{PW99}.   
Reverberation mapping also demonstrated a strong anti-correlation between the line width and the time delays of the line 
response to the continuum variations \cite{PW00}. 
The BLR size, as measured from the \mbox{C\,{\sc iv}}~1549{\AA} line delays, 
scales with the accretion luminosity as $R_{\ \mbox{\scriptsize C\,{\sc iv}}} \approx 0.2 L_{47}^{1/2}$ pc \cite{Kaspi07}. 
However, the delays in the high-ionization He\,{\sc ii}~1640{\AA} line as observed in Seyferts
are 3--5 times smaller  \cite{Korista95,PW99}. On the other hand, the Balmer lines  give sizes 2--3 times larger. 
It is also worth mentioning that the scaling \cite{Kaspi07} extrapolated to the high-luminosity quasars is based on one object 
(S5 0836+71) and one line (\mbox{C\,{\sc iv}}~1549{\AA}), while the BLR size estimated 
in the Seyfert-luminosity range has more than an order of magnitude spread. Thus the actual BLR size in quasars 
is highly uncertain. 
The strongest BLR line, Ly$\alpha$, does not show any variability on the timescale of years
in the four quasars in which it was observed \cite{Kaspi07}, 
implying that it is produced much further away that \mbox{C\,{\sc iv}}~1549{\AA}. 
This has important implications for the opacity of the GeV photons. 

\begin{figure}
\centerline{\epsfig{file= 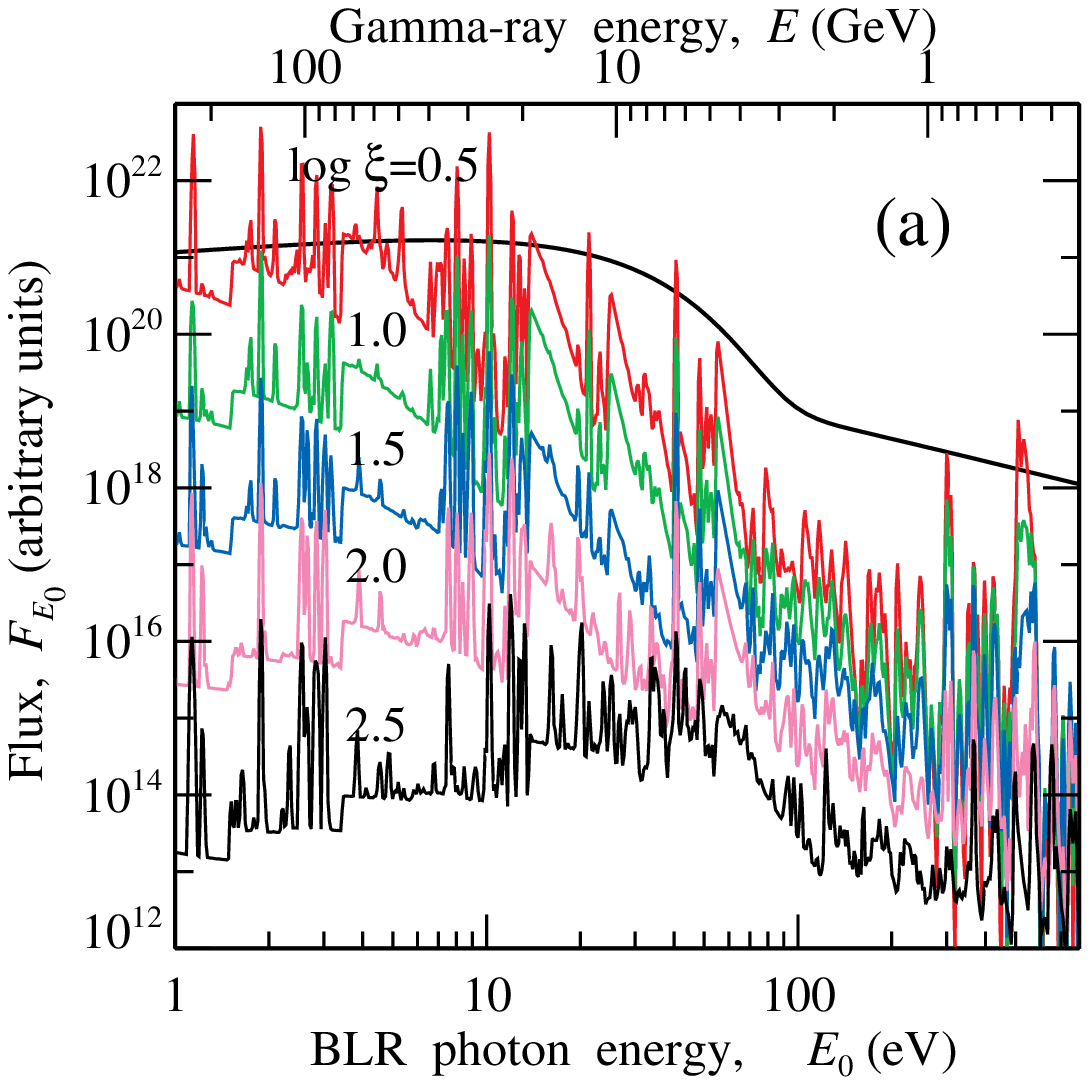,width=7.3cm}
\hspace{0.3cm}
\epsfig{file= 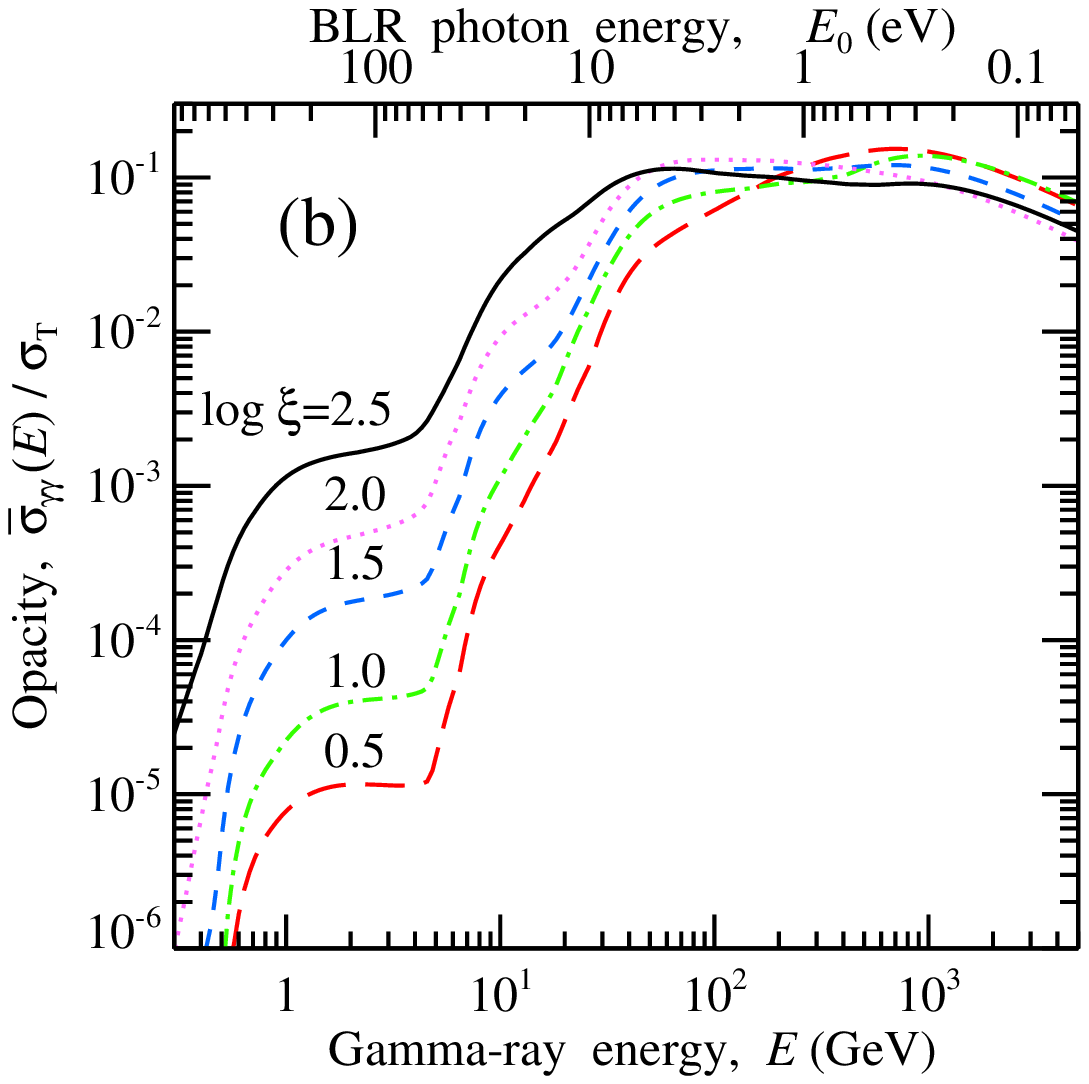,width=7.3cm}} 
\caption{ (a) Spectra $F_E=d N/d\ln E$ 
of the BLR clouds of column density $N_{\rm H}=10^{23}$ cm$^{-2}$ 
and various  ionization parameters $\xi$ photoionized by a quasar continuum shown by a smooth curve and consisting 
of a multicolor disc with the maximum temperature $10^5$ K and 
a power law tail of photon index $\Gamma=2$ extending to 100 keV with 10\% of the total luminosity.
We use photoionization code {\sc xstar} vs. 2.2 \cite{KB01}. 
The upper axis show the threshold energy for $\gamma\gamma$ reaction 
 corresponding to the BLR photon, given by Eq. (\protect\ref{eq:threshold}). 
(b) Angle-average cross section (in units of Thomson cross section) for $\gamma$-rays due to $\gamma\gamma$
absorption on the BLR photons with the spectra shown in panel (a).  
Jumps in opacity are clearly seen at energies corresponding to the strongest lines and recombination continua. 
The 0.3--0.5 GeV jump for large $\xi$ is produced by O\,{\sc viii} 16--19{\AA\AA}  lines, the jump at low $\xi$ is due to the 
O\,{\sc vii}  22{\AA} line complex. The jump at $\sim$5 GeV visible at all ionizations 
is due to the He\,{\sc ii} LyC and Ly lines at 40--60 eV. 
The 20--30 GeV jump seen at low $\xi$ is produced by H\,{\sc i} Ly lines and continuum and the C\,{\sc iv} 1549{\AA} line.  
From \cite{PS10}.
}	
\label{fig:blr}
\end{figure}

Using  {\sc xstar} we   generated \cite{PS10} a grid of photoionization models
of the BLR clouds assumed to be simple slabs of constant gas density  $n_{\rm H}$, 
of fixed column density  at $N_{\rm H}=10^{23}$ cm$^{-2}$, and a clear view to the ionizing source. 
We varied the ionization parameter  $\xi=L/(r^2 n_{\rm H})$  from 10$^{0.5}$ to 10$^{2.5}$. 
The resulting spectra are shown in Fig.~\ref{fig:blr}a. 
BLR radiation consists mostly of lines and recombination continua, which 
in the UV range produce narrow, line-like features, because  the temperature of the photoionized regions is typically 
much below the corresponding ionization potentials of H and He.

If the $\gamma$-rays are produced within the BLR, they propagate through a roughly isotropic photon field.
The absorption strength  by the isotropic line photons of energy $E_0$ 
can be characterized by the Thomson optical depth \cite{PS10}: 
\begin{equation} \label{eq:taut}
\tau_{\rm T} 
=  N_{\rm ph}  \sigma_{\rm T} \approx 
 \frac{L_{\rm line}}{4\pi R^2 c E_0}  R \sigma_{\rm T}
\approx 110 \ \frac{L_{\rm line,45} }{R_{18}} \frac{10\ {\rm eV}}{E_0} ,
 \end{equation} 
where $N_{\rm ph}$ is the column density of line photons along the line of sight,  $\sigma_{\rm T}$ is the Thomson cross-section, 
$L_{\rm line}$ is the line luminosity, and $R$ is the typical size. (Here  we define $Q=10^xQ_x$ in cgs units.)
The energy-dependent opacity is 
\begin{equation} \label{eq:tauEE}
\tau_{\gamma\gamma}(E,E_0) = N_{\rm ph}  \sigma_{\gamma\gamma}(s) 
= \tau_{\rm T} \frac{\sigma_{\gamma\gamma}(s)}{\sigma_{\rm T} }. 
 \end{equation} 
Here $\sigma_{\gamma\gamma}(s)$ is the angle-averaged $\gamma\gamma$ cross-section, which 
has a threshold at $s\equiv EE_0/(m_{\rm e}c^2)^2=1$, i.e.
\begin{equation} \label{eq:threshold}
E_{\rm th}=261\ {\rm GeV}/(E_0 [{\rm eV}]), 
 \end{equation} 
rapidly grows to the maximum of $\sim$0.21$\sigma_{\rm T}$ at the energy of $\sim$3.5 times the threshold, 
and then slowly decreases roughly inversely with energy  \cite{GS67,ZDZ88,Aha04}.   
High-energy photons transmitted through the soft line photons 
are attenuated as $\propto \exp(-\tau_{\gamma\gamma}(E,E_0))$.
Approximating the spectrum by a power law,  we can estimate the break in the power-law index 
at $E_{\rm th}$ produced by an individual strong line (or recombination continuum) as
\begin{equation} \label{eq:deltagamma}
\Delta\Gamma=- \frac{{\rm d} \ln  \exp(-\tau_{\gamma\gamma}(E,E_0))}{{\rm d} \ln E} \approx  
\tau_{\rm T} \max \frac{\sigma_{\gamma\gamma}(s)/\sigma_{\rm T}}{\ln s } \approx \frac{\tau_{\rm T} } {4}.
\end{equation}

Equation (\ref{eq:taut}) shows that in powerful quasars with the 
Ly$\alpha$ luminosity of about $10^{45}$~erg~s$^{-1}$, the $\gamma$-rays above 30 GeV have troubles of escaping 
unless the density of the line photons is sufficiently small (e.g. if they are produces in a large volume of $R\gtrsim3$ pc), 
or if the $\gamma$-rays are produced outside the BLR. 
As the BLR size $R$ (in a given line) scales with the total luminosity $L^{1/2}$ \cite{Kaspi07}, 
the opacity also scales similarly (assuming a constant covering factor of BLR clouds)
\begin{equation} 
\tau_{\rm T} \propto L^{1/2},
\end{equation}
and obviously the $\gamma\gamma$ absorption can be important only in the high-luminosity sources such as FSRQs. 
Another important consequence of Eq. (\ref{eq:taut})  is that   it is not the luminosity, 
but $L/R$ ratio (i.e. compactness) that determines the role of a specific line in absorption of the $\gamma$-rays.

Because of strong stratification of the BLR extending over two orders of magnitude in distance \cite{Krolik99,AGN2}, 
any model assuming that all lines are produced 
at the same distance from the central source \cite{Liu06,Reimer07,TM09} strongly underestimates the 
$\gamma\gamma$-opacity by the relatively weak, high-ionization lines produced in a small volume
(if the $\gamma$-rays are produced closer to the centre)
and overestimates the opacity for multi-GeV photons from the strong, low-ionization lines which 
are produced  in a larger volume. 

The total $\gamma\gamma$-opacity associated with a given BLR shell at a fixed ionization  
can be determined by a convolution of  the opacity for the line radiation
with the BLR photon column density spectral distribution  
\begin{equation}
\overline{\tau}_{\gamma\gamma} (E) =  N_{\rm ph}  \overline{\sigma}_{\gamma\gamma}(E) =
 \tau_{\rm T}  \frac{\overline{\sigma}_{\gamma\gamma} (E)}{\sigma_{\rm T}}  = 
\int \sigma_{\gamma\gamma}  (s) \frac{d N_{\rm ph}}{d E_0} {\rm d} E_0 .  
\end{equation}
The  $\gamma\gamma$  cross sections weighted with the BLR spectra from Fig.~\ref{fig:blr}a are shown 
in Fig.~\ref{fig:blr}b. 
If ionization is high, the dominant photon source is the He\,{\sc ii} complex at 40--60 eV, which  produces a break 
in opacity at 4--7 GeV. Above 10 GeV, the opacity is a rather smooth function of energy with additional absorption coming 
from He\,{\sc i}, hydrogen Ly lines and recombination continuum, and C\,{\sc v} 2274{\AA}. In the 0.3--0.7 GeV region the opacity 
has another break due to high-ionization lines of O\,{\sc vii}.   
In the low-ionization environment, the strongest absorption is produced by hydrogen Ly lines and LyC, 
forcing a break in opacity at about 20--30 GeV. At the same time, the break due to helium is also clearly visible at a few GeV. 
The opacity is nearly flat up to 1 TeV because of the contribution from additional lines: 
Mg\,{\sc ii}  2800{\AA}, H$\alpha$, and He\,{\sc i} 10832{\AA}.
As we argued above, the complete model for the opacity has to be necessarily multi-zone and 
 include radial stratification of the BLR and 
 probably the angular dependence of the BLR radiation, which can affect the threshold energies.

\section{Blazars and absorption by \mbox{He\,{\sc ii}} and \mbox{H\,{\sc i}}  LyC}

For the first analysis \cite{PS10}, we have selected  several brightest FSRQs from the sample of 12 objects in Table 1 of  \cite{Abdo10_blazars} and  have chosen the same 180-day interval for easier comparison.
We have developed our own software for data analysis and compared  its results for the simple 
broken power law models with those obtained by the standard maximum likelihood analysis tool {\sc gtlike} 
in \cite{Abdo10_blazars}. 
For a few objects the fits with the power-law are acceptable, while other 
show clear breaks (see Fig. \ref{fig:spectra}a) that are rather well described by  
 the broken power law model (see details in \cite{PS10}). 
 
To check the hypothesis that the BLR photons are responsible for the breaks, we have applied a simplified 
model of the BLR consisting of two strongest features:  \mbox{He\,{\sc ii}} LyC  at 54.4 eV and \mbox{H\,{\sc i}} LyC
at 13.6 eV. We call this double-absorber (DA) model.  The free parameters are the optical depths $\tau_{\rm He}$ and $\tau_{\rm H}$ in these``lines''.  
The observed absorption threshold were redshifted by the appropriate $1+z$ factor. 
Assuming that the underlying continuum is a power-law, we obtained a good fit (see Fig. \ref{fig:spectra}a). 
In RGB J0920+446 (with $z= 2.19$), 
the absorption is seen only at high energies with the break energy 
corresponding to the pair-production threshold on \mbox{H\,{\sc i}} LyC, 19.2 GeV$/(1+z)\sim 6$ GeV,
while in other cases the first break is close to the threshold for the absorption on \mbox{He\,{\sc ii}} LyC, 4.8 GeV$/(1+z)$,
and there are indication of the second break at higher energy due to  \mbox{H\,{\sc i}} LyC.

\begin{figure}
\centerline{\epsfig{file=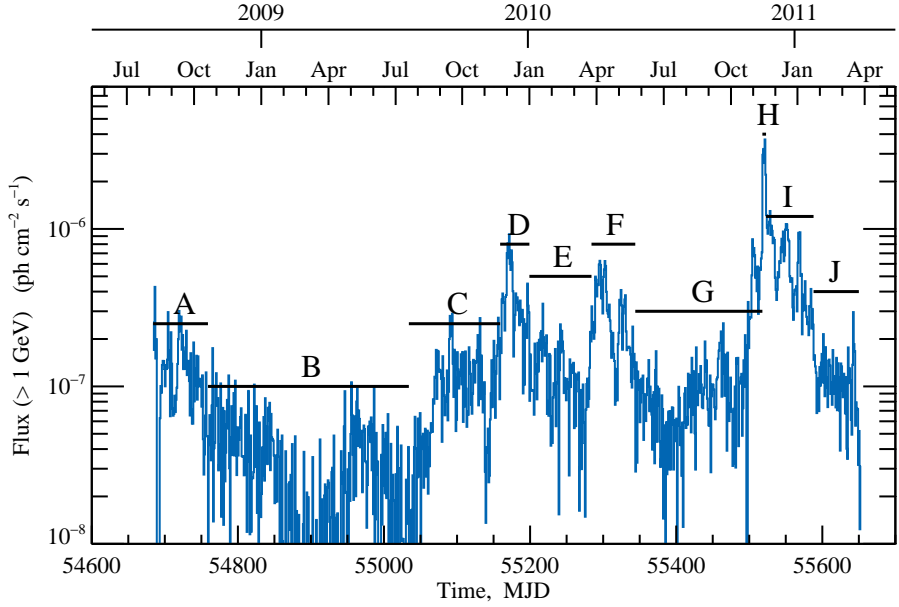,width=12cm}} 
\caption{The light curve of FSRQ 3C 454.3 above 1 GeV as observed by {\it Fermi}/LAT. Adapted from \cite{SP11}. 
}\label{fig:lc}
\end{figure}

The FSRQ  3C 454.3 at redshift $z=0.859$ is by far the brightest blazar in the $\gamma$-ray range during the lifetime  
of {\em Fermi}. 
The source shows variations in brightness by two orders of magnitude (see Fig. \ref{fig:lc}).
LAT has detected more than hundred thousand photons from this source. 
With such a rich photon statistics, it is worth trying to study the spectral variation at various flux levels, 
to constrain the shape of the underlying spectrum,  and to search for correlations between flux and 
the $\gamma\gamma$ absorption depth. The last aim is the most important one, as it allows to learn about 
variations in the location of the $\gamma$-ray emitting region.

We have split the data into the ten time intervals covering periods of more or less constant flux levels  
and analyzed the spectrum in each interval. The spectra for the high- (interval H) and low-flux (B) intervals as well as
the average spectrum are shown in Fig. \ref{fig:spectra}b.
Clear deviations from a simple power-law spectrum are obvious, and addition of 
the DA still does not  provide a good fit at all flux levels. Thus instead of the power-law 
to describe the underlying continuum we have tried a lognormal 
function $EF_E \propto 10^{-\log^2 (E/ E_{\rm peak}) /\sigma_{\rm ln}^2}$ plus the DA. 
The fits with this model are superior compared to the previously used simple phenomenological models,  
especially at high fluxes with high photon statistics.
The data can be also well fitted with $\gamma\gamma$ absorption produced by the full BLR spectrum 
of high-ionization (see dashed line in Fig. \ref{fig:spectra}b). 
A large contribution to the $\chi^2$ of the spectral fits comes actually from the high-energy tail.
This might be an evidence for variations of the absorber optical depth during the observation 
and the fact that the sum of the spectra with low and high opacities differs from the spectrum at a 
fixed intermediate opacity. 
A probable excess of  photons above 15 GeV (especially for the average spectrum) 
can also indicate that the spectrum is a superposition of emission states with different opacities 
due to the extended or moving emission zone. 
This proposal is supported by the spectral variability seen 
during the latest flare in 2010 November \cite{Fermi11_3C454}, 
where photons above 10 GeV arrive mostly at the end of the flare, clearly 
indicating the position of the $\gamma$-ray emitting region further away from the black hole.

\begin{figure}
\centerline{\epsfig{file= 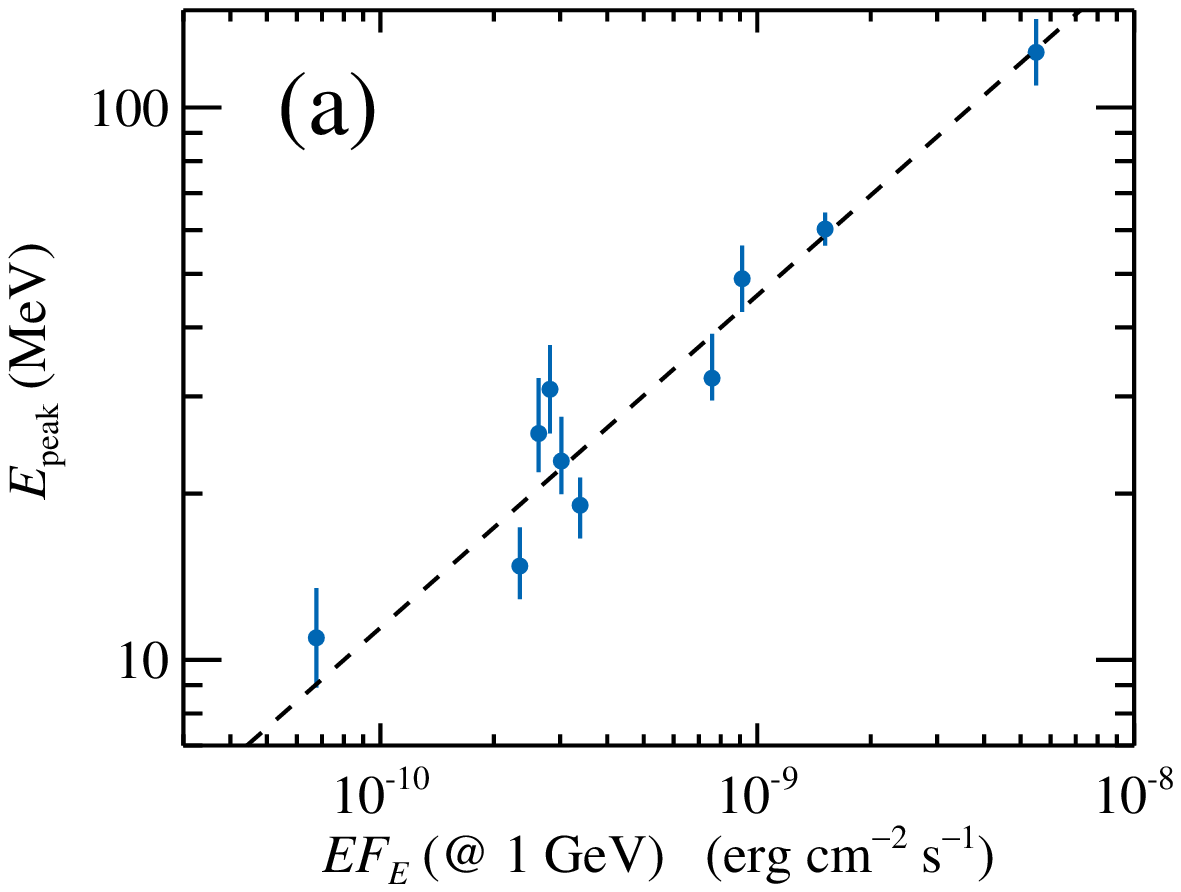,width=7.5cm} 
%\hspace{1cm}
\epsfig{file=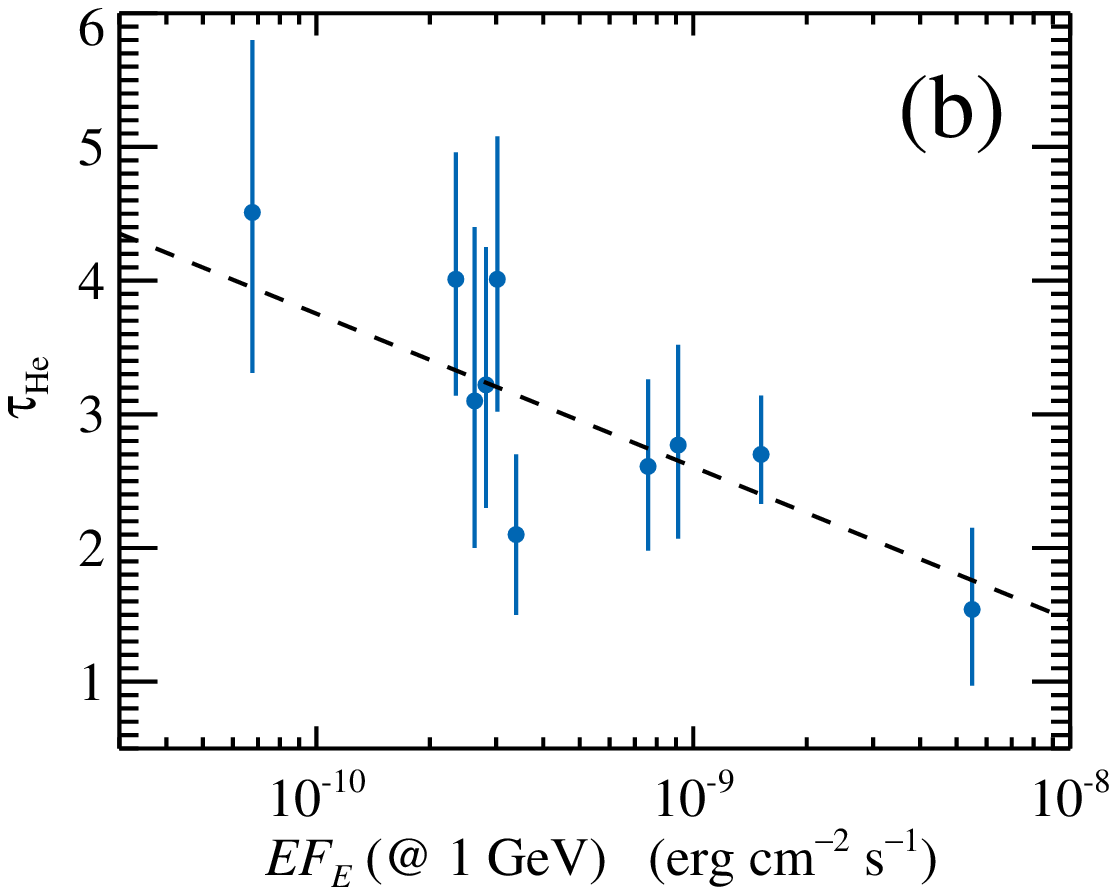,width=7.5cm} } 
\caption{(a) Dependence of the peak (in $EF_E$) of the best-fitting lognormal  
(with $\sigma_{\rm ln}=2.7$) + DA model on flux at 1 GeV as observed by {\it Fermi}/LAT from 3C 454.3.   
(b) Dependence of the Thomson optical depth in the \mbox{He\,{\sc ii}} LyC 
(defined by Equation (\protect\ref{eq:taut})) on flux at 1 GeV. 
The dashed line shows the best-fitting relation given by Equation (\protect\ref{eq:tauHe_fit}).
Adapted from \cite{SP11}. 
}\label{fig:epeak_tauHe}
\end{figure}

3C 454.3 demonstrates a highly significant  correlation of the peak energy $E_{\rm peak}$ with flux 
(similar to the hardness--flux correlation  \cite{Fermi10_3C454}), 
which is well represented by a power-law (Fig. \ref{fig:epeak_tauHe}a): 
\begin{equation} \label{eq:epeak_fit}
E_{\rm peak} = (46\pm2) \ \mbox{MeV}\ F_{-9}^{0.60\pm0.04},
\end{equation}
where $F=EF_E$ at 1 GeV. The opacity $\tau_{\rm H}\sim3$--7 did not vary significantly, but 
we observed an anti-correlation between $\tau_{\rm He}$ and the flux  (see Fig. \ref{fig:epeak_tauHe}b):
\begin{equation} \label{eq:tauHe_fit}
 \tau_{\rm He} = (2.61\pm0.22) - (1.15\pm 0.47)\ \log  F_{-9}.
\end{equation}

The strength of absorption at GeV energies together with the estimations of the 
luminosities in LyC in principle constraints the BLR size (from Eq. \ref{eq:taut}), where most of 
this soft radiation is produced. Thus, we get: 
\begin{equation} 
R_{\rm He} \approx 0.22\ L_{\rm He,44}  (\tau_{\rm He}/3) \ \mbox{pc} , \quad 
R_{\rm H} \approx 3.7\ L_{\rm H,45}  (\tau_{\rm H}/7) \ \mbox{pc} .
\end{equation}
The luminosities of 3C 454.3 in Ly$\alpha$ and He\,{\sc ii}~1640{\AA} are 
$\sim10^{45}$~erg~s$^{-1}$  and $6\times10^{43}$~erg~s$^{-1}$, respectively \cite{Wills95}. 
Assuming equal luminosities in LyC and  Ly$\alpha$, the resulting 
$R_{\rm H}$ is much larger than the usually quoted sub-pc size of BLR. It is, however, 
consistent with the absence of Ly$\alpha$ variability in powerful quasars \cite{Kaspi07}. 
The He\,{\sc ii} LyC luminosity is probably larger than $10^{44}$~erg~s$^{-1}$,
but is model dependent and therefore the estimate of $R_{\rm He}$ is less certain. 

A rather large ratio $\tau_{\rm He}/\tau_{\rm H}$ (between about 1/4 and 1) 
and the observed $\tau_{\rm He}$--$F$ anti-correlation 
indicate that the $\gamma$-ray emission region lies close to the boundary $R_{\rm He}$ 
of the complete ionization of helium and moves out at higher luminosity. 
A small distance  to the beginning of the $\gamma$-ray emitting region 
alleviates the problems created by observations of the fast variability from 3C 454.3 \cite{TGB10,Bonnoli11}.

\section{Absorption features in the  CTA range}

\begin{figure}
\centerline{\epsfig{file= 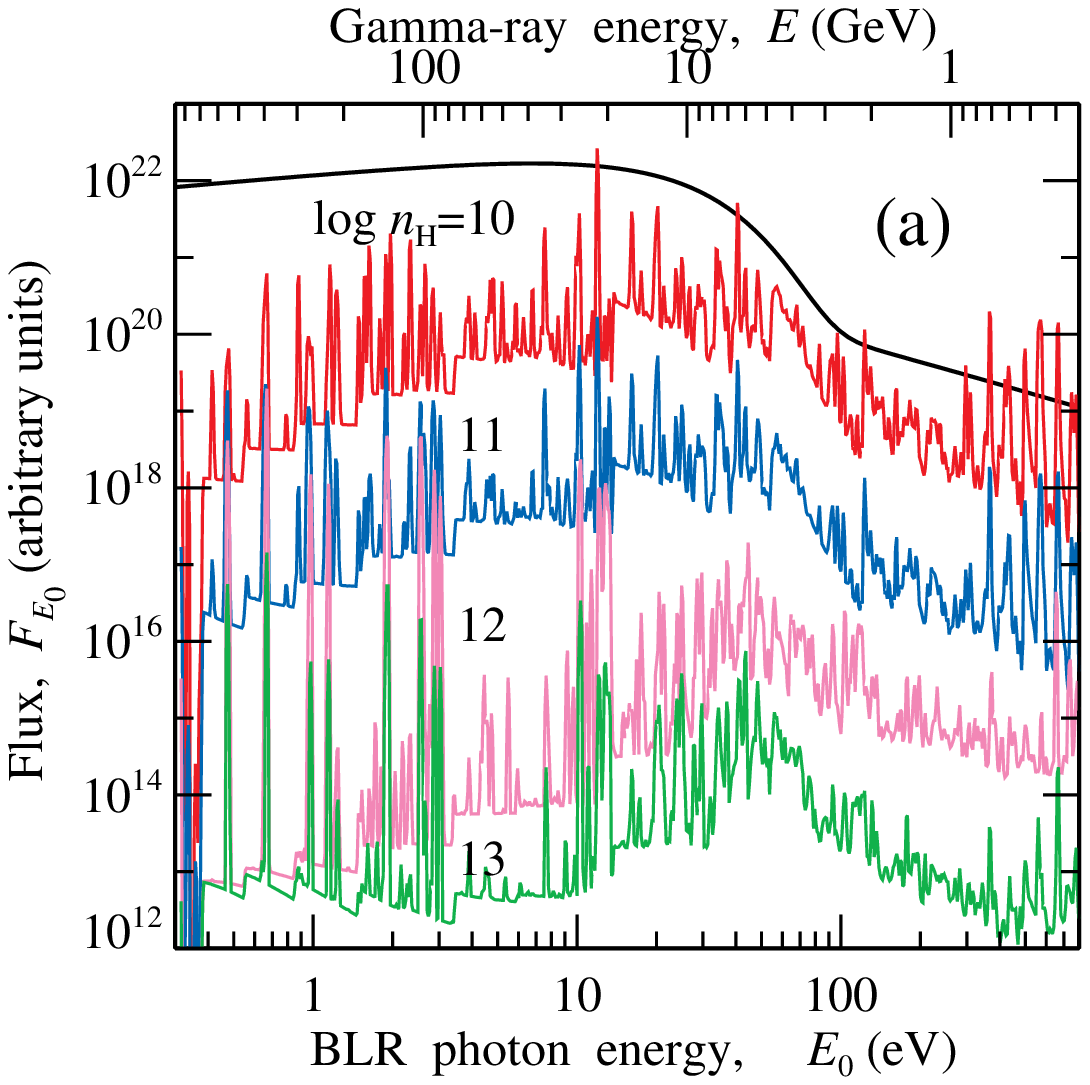,width=7.3cm}
\hspace{0.3cm}
\epsfig{file= 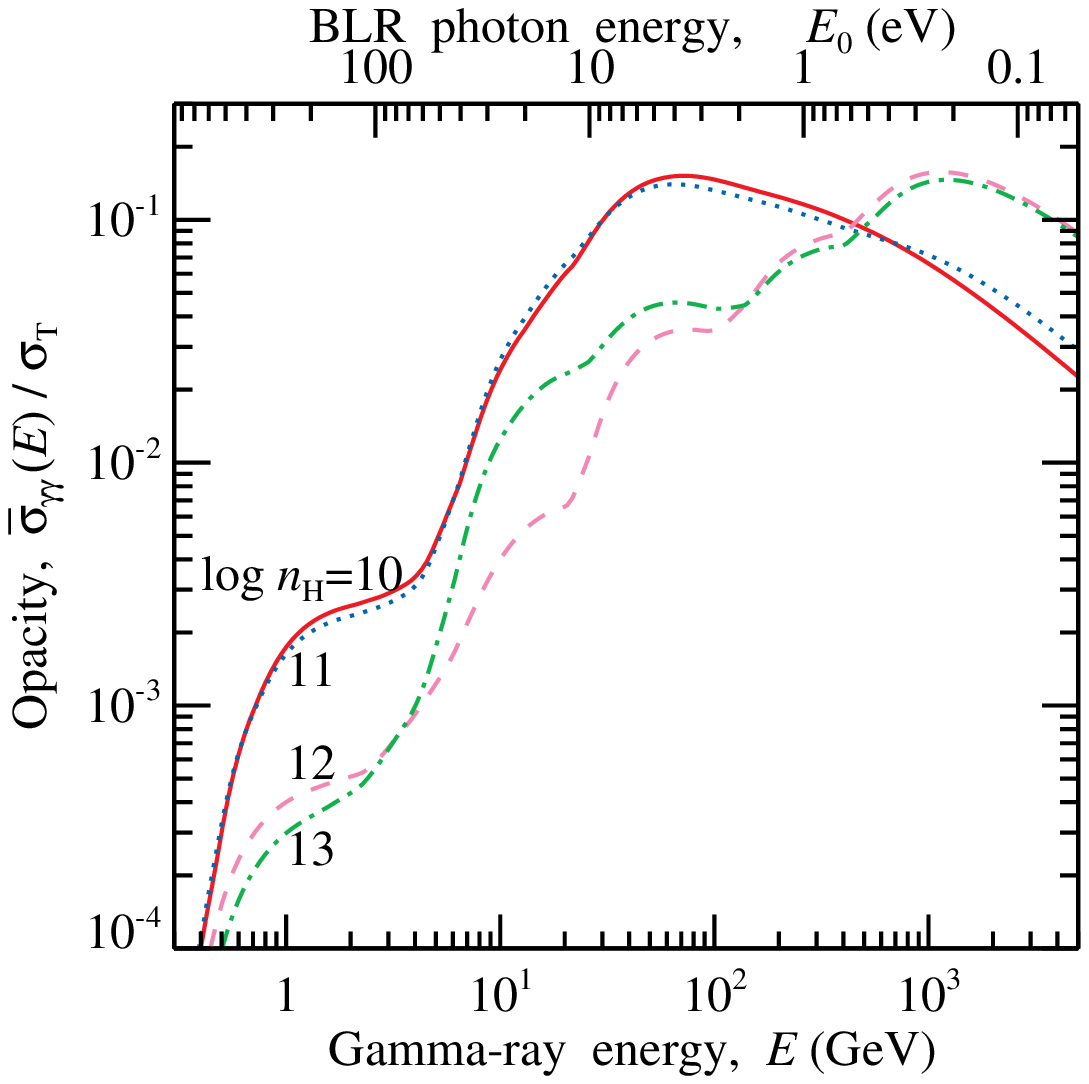,width=7.3cm}} 
\caption{(a) Same as Fig. \protect\ref{fig:blr}a, but here 
we fix the ionization parameter at $\log\xi=2.5$ and vary the density of the clouds 
from $n_{\rm H}=10^{10}$ (at the top) to $10^{13}$ cm$^{-3}$ (at the bottom) with steps of a factor of 10. 
(b) Corresponding $\gamma\gamma$ cross sections weighted with the BLR spectrum are shown
by the solid, dotted, dashed and dot-dashed curves, respectively. 
In the high-density regime there are additional jumps in opacity 
at $\sim$100 GeV due to the Balmer lines and at $\sim$400 GeV due to the Pa$\alpha$ line. }	
\label{fig:blrdens}
\end{figure}

As we have shown in Fig. \ref{fig:blr}, the strongest spectral features responsible 
for the $\gamma\gamma$ absorption and the breaks in the {\em Fermi}/LAT band are LyC of hydrogen and ionized helium. 
At energies above 30 GeV, there are no significant spectral features expected, just because 
there are many closely-spaced BLR lines that produce rather smooth absorption profile.
However, these simulations only covered a rather narrow set of parameters. 
The BLR not only has a large spread in the ionization parameter, but  the cloud density may also 
vary by orders of magnitude.  For simplicity we fixed $\log\xi$ and  looked at the effects of the varying density
(see Fig. \ref{fig:blrdens}). We see that in the high-density  regime, the BLR spectrum below a few eV 
is dominated by lines of the hydrogen series. Strong Balmer lines (H$\alpha$, H$\beta$) 
give a rise to the $\gamma\gamma$ opacity at 100--140 GeV and the Pa$\alpha$ line is responsible 
for another rise at $\sim$400 GeV. Thus spectral breaks at these energies are expected if multi-GeV photons 
are transmitted through such a BLR. Detection of such breaks with the Cherenkov Telescope Array (CTA) will give a clue to the structure 
and physical conditions within the BLR.

\section{Conclusions}

We have shown that the spectral breaks detected by the {\em Fermi}/LAT in powerful blazars at a few GeV 
can be naturally explained by the $\gamma\gamma$ absorption on the hydrogen and  \mbox{He\,{\sc ii}} LyC.
The strength of the absorption ultimately proves that the blazar zone has to be located within or close to the 
high-ionization zone of the BLR. Variability in the column density of  \mbox{He\,{\sc ii}} LyC 
that produces absorption above $\sim$3 GeV in the brightest GeV blazar 3C~454.3 implies 
that the $\gamma$-ray emitting region is located close to the boundary of fully ionized He zone and is 
moving away from the black hole when the flux increases.
The strength of the Ly$\alpha$ in this object allows to estimate the BLR size in LyC,
which seems to exceed significantly the estimations previously discussed in the literature. 
A relatively large BLR size dilutes the density of the BLR photons and allows the multi-GeV photons to escape. 
Detections of a few powerful blazars in the TeV range thus does not necessarily means that the 
$\gamma$-ray emitting region is located outside the BLR as argued recently \cite{Magic11_1222,Tanaka11,TBG11}. 
On the contrary, the GeV breaks are impossible to produce far away from the BLR as there are 
no enough photons available for absorption and there is no any other physical mechanism 
that can produce sharp spectral breaks. 
High-density clouds embedded within the BLR can produce strong lines of hydrogen series which 
can absorb multi-GeV photons. Detection of the spectral breaks at $\sim$100 and $\sim$400 GeV with the CTA
will put interesting constraints on the BLR physical conditions and the location of the $\gamma$-ray emission 
zone.

\section*{Acknowledgments}
This research was supported  by the Academy of Finland grant 127512  and the Magnus Ehrnrooth foundation.
The research made use of public data obtained from the Fermi Science Support Center. 
We thank Maxim  Barkov, Markus Boettcher, Evgeny Derishev,  Chuck Dermer, Gabriele Ghisellini,
Julian Krolik, Amir Levinson, Fabricio Tavecchio, and Dmitry Yakovlev  for useful discussions 
 and Tim Kallman  for his help with {\sc xstar}.

\end{document}